\documentclass[conference]{IEEEtran}

\usepackage{cite}

\ifCLASSINFOpdf
   \usepackage[pdftex]{graphicx}
\else
\fi
\usepackage[cmex10]{amsmath}
\usepackage{array}

\usepackage[table]{xcolor}

\begin{document}
%
\title{Circuit-Level Evaluation of the Generation of \\ Truly Random Bits with  Superparamagnetic \\ Tunnel Junctions}



%
\author{\IEEEauthorblockN{Damir Vodenicarevic\IEEEauthorrefmark{1},
Nicolas Locatelli\IEEEauthorrefmark{1},
Alice Mizrahi\IEEEauthorrefmark{1}\IEEEauthorrefmark{3}, 
Tifenn Hirtzlin\IEEEauthorrefmark{1}, \\
Joseph S. Friedman\IEEEauthorrefmark{2},
Julie Grollier\IEEEauthorrefmark{3} and
Damien Querlioz\IEEEauthorrefmark{1}}
\IEEEauthorblockA{\IEEEauthorrefmark{1}Centre for Nanoscience and Nanotechnology, CNRS, Univ Paris-Sud, Universit\'e Paris-Saclay\\
rue Andr\'e Amp\`ere, 91405 Orsay, France
\\ Email: damien.querlioz@u-psud.fr}
\IEEEauthorblockA{\IEEEauthorrefmark{2}University of Texas at Dallas, USA}
\IEEEauthorblockA{\IEEEauthorrefmark{3}UMP CNRS/Thales, Univ Paris-Sud, Universit\'e Paris-Saclay, Palaiseau, France}
}

\maketitle

\begin{abstract}
Many emerging alternative models of computation require massive numbers of random bits, but their generation at low energy is currently a challenge. The superparamagnetic tunnel junction, a spintronic device based on the same technology as spin torque magnetoresistive random access memory has recently been proposed as a solution, as this device naturally switches between two easy to measure resistance states, due only to thermal noise. Reading the state of the junction naturally provides random bits, without the need of write operations. In this work, we evaluate a circuit solution for reading the state of superparamagnetic tunnel junction. We see that the circuit may induce a small read disturb effect for scaled superparamagnetic tunnel junctions, but this effect is naturally corrected in the whitening process needed to ensure the quality of the generated random bits. These results suggest that superparamagnetic tunnel junctions could generate truly random bits at 20~fJ/bit, including overheads, orders of magnitudes below CMOS-based solutions.
\end{abstract}


%
\IEEEpeerreviewmaketitle

\section{Introduction}

With the end of Moore's law in sight, microelectronics research is currently exploring multiple alternative directions for replacing or complementing current architectures based on the von Neumann principles.
Several of these ideas explore the power of random numbers for computing. In particular, stochastic computing approaches can achieve approximate computations with  minimal circuit overhead \cite{alaghi2013survey,friedman_bayesian_2016}. Several neural network paradigms also exploit high quantities of random numbers \cite{merolla_million_2014,suri2013bio,esser2013cognitive}, mirroring an idea in neuroscience that our brain might be exploiting noise \cite{hamilton2014stochastic}. Finally, random logic gates might also allow reversible forms of computation \cite{camsari2017stochastic}.

However, the generation of massive amounts of random bits at low energy is currently a challenge that limits the applicability of such ideas.
In recent years, emerging nanodevices have been introduced as a remarkable alternative for random bit generation \cite{ balatti2015true }. In particular, several studies have reported that the same magnetic tunnel junctions used as the basic cells of spin torque magnetoresistive memory (ST-MRAM) can also be used as random bit generators: when they are programmed with short programming states, the final state of the junction is probabilistic \cite{fukushima_spin_2014,choi_magnetic_2014,fong_generating_2014,lee2017design, qu2017true}. 
Unfortunately, programming these junctions consumes significant energy, in the picoJoules range.
In Ref.~\cite{PRA}, we showed experimentally that superparamagnetic tunnel junctions, which are variations of the magnetic tunnel junctions of ST-MRAMs, can produce high quality random bits using much lower energy consumption, possibly providing an ideal device for randomness-based alternative forms of computing.
These devices indeed switch between their two stable states due to random noise only: 
just reading the state of the device provides random bit, without the need of write operations.

However, due to their physics, these devices might be disturbed by the readout circuitry.
In this paper, we investigate, through physical modeling and circuit simulation, the sensitivity of superparamagnetic tunnel junctions to read disturb effects, and conclude that it does not constitute a concern for the technology if the readout circuit is properly chosen.
We first introduce the basic principles of superparamagnetic tunnel junctions. We then present the modeling of read disturb effects in these devices, and the results of our circuit-level investigation.

\section{Random Bit Generation with Superparamagnetic Tunnel Junctions}

\begin{figure}[!t]
\centering
\includegraphics[width=\columnwidth]{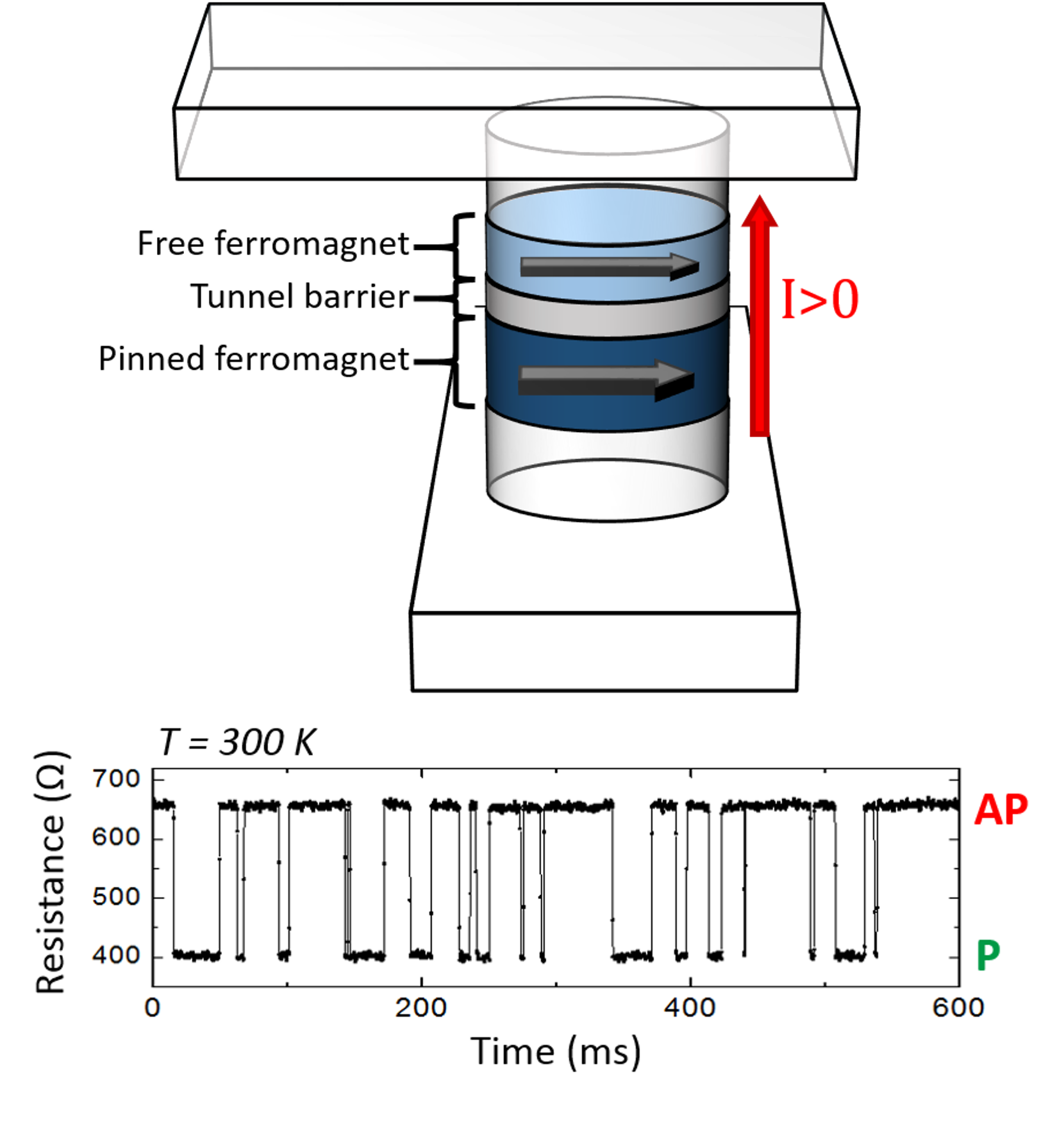}
\caption{(a) Simplified illustration of a typical superparamagnetic tunnel junction. (b) Example measurement of the electrical resistance of a superparamagnetic tunnel junction as a function of time.}
\label{fig:schemaMTJ}
\end{figure}

Superparamagnetic tunnel junctions are nanodevices composed of several thin layers of magnetic and non-magnetic materials (Fig.~\ref{fig:schemaMTJ}(a)). In particular, they feature two nanomagnets: a pinned nanomagnet and a free one. The magnetization of the free nanomagnet can be anti-parallel (AP) or parallel (P) to the magnetization of the pinned nanomagnet.  Superparamagnetic tunnel junctions therefore possess two states  that differ in the electrical resistance of the device. The discernibility of the two resistance states is characterized by the tunnel magnetoresistance~(TMR) parameter:
\begin{equation}
TMR = \frac{R_{AP}-R_P}{R_P}.
\label{TMR}
\end{equation} 

Superparamagnetic tunnel junctions are based on the same technology as ST-MRAM. In a ST-MRAM cell, the AP and P states are highly stable, leading to a non-volatile memory behavior. By contrast, in a superparamagnetic tunnel junction, the two states are intentionally unstable: the junction naturally switches between its two states due to thermal energy only\cite{locatelli2014noise, mizrahi2016controlling,mizrahi2015magnetic}. Fig.~\ref{fig:schemaMTJ}(b) presents a brief experimental measurement of the resistance a superparamagnetic tunnel junction as a function of time. Such a device is extremely attractive for random bit generation as it naturally transforms a weak but truly random effect (thermal noise) into a macroscopic and easy-to-extract signal (the electrical resistance of the junction).

\begin{figure}[!t]
\centering
\includegraphics[width=\columnwidth]{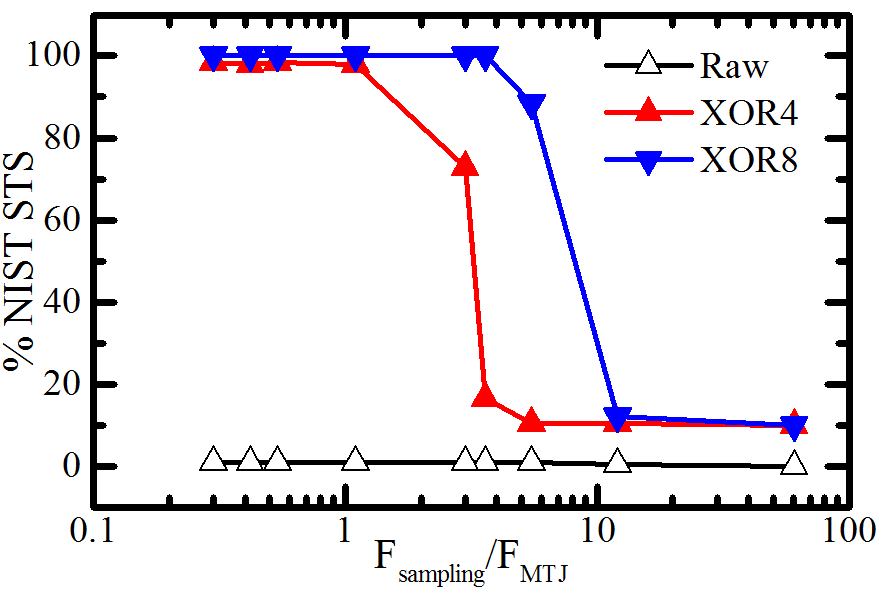}
\caption{Experimental results based on several days of measurements on superparamagnetic tunnel junctions (based on Ref.~\cite{PRA}). Proportion of NIST STS tests compatible with cryptographic quality in the absence of whitening (Raw) or with XOR-based whitening (XOR4 and XOR8), for different sampling rates. $\mathrm{F}_\mathrm{MTJ}$ is the natural mean frequency of the junction.}
\label{fig:NIST}
\end{figure}

Such junctions can generate random bits of high quality. Long sequences of random bits were obtained by measuring junctions for days, providing gigabits of data, at different sampling frequencies. When a junction is measured, the output is said to be one if the junction was in the AP state, and zero if was in the P state. The obtained sequences of bits were tested using the National Institute of Standards and Technology Statistical Test Suite (NIST STS) for random bit quality \cite{soto_statistical_1999}, either directly (``Raw'') or after a simple ``whitening'' process. Whitening aims at eliminating biases in the random bits and consists in combining four (``XOR4'') or eight (``XOR8'') independent bitstreams into a single one through XOR gates. The NIST STS separates the resulting bitstream into 1~Mbit sequences and tests them against 188 different tests. Fig.~\ref{fig:NIST}, based on the experimentally obtained bitstreams of Ref.~\cite{PRA}, shows the proportion of NIST tests satisfying cryptographic quality criteria as a function of the sampling rate, and for different whitening processes. 

The sequences whitened with the XOR8 process and sampled below $\mathrm{F}_\mathrm{sampling}/\mathrm{F}_\mathrm{MTJ} = 3.6$ pass all the tests, therefore demonstrating high random bit quality. By contrast, Raw sequences do not pass the tests, due to residual stray fields in the nanodevices that cause their mean values to deviate slightly from 0.5. Sequences that are sampled too fast also do not pass the tests, as they exhibit auto-correlation.

In Fig.~\ref{fig:NIST}, sampling frequencies are expressed relatively to the measured mean frequency of the devices $F_\mathrm{MTJ}$, which is $1.4kHz$ in the measured devices. 
The results of Ref.~\cite{PRA} show that $F_\mathrm{MTJ}$ scales with device area, and that scaled superparamagnetic tunnel junctions might generate random bits at  frequencies higher than $MHz$. 
They also highlight important design rules in terms of cross talk and sensitivity to environmental perturbations (temperature, magnetic field).
We now evaluate the impact of the readout operation on the generation of random bits.

\begin{figure}[!t]
\centering
\includegraphics[width=2.5in]{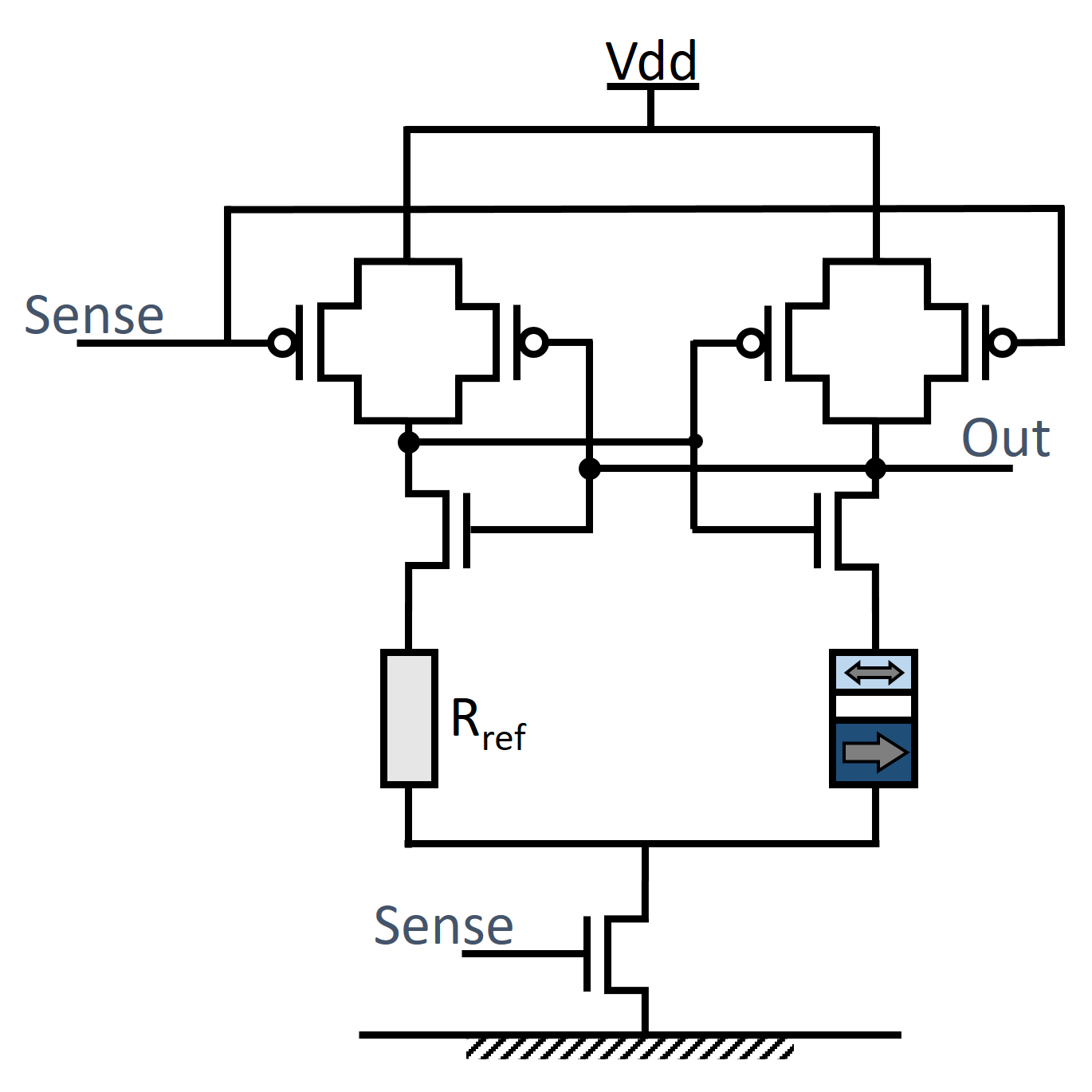}
\caption{Schematic of a precharge sense amplifier circuit \cite{zhao_high_2009}. $R_\mathrm{ref}$ is a reference resistance whose value is chosen to lie between the resistances of the P state ($R_P$) and the AP state ($R_{AP}$) of the superparamagnetic tunnel junction.}
\label{fig:PCSA}
\end{figure}

\section{Modeling of Read Disturb Effects}

Electrical currents influence magnetic tunnel junctions through the spin transfer torque physical effect. High spin transfer torque is highly desirable for memory applications of magnetic tunnel junctions, as it allows their efficient programming. By contrast, in our random bit generation example, it acts as a perturbation, possibly affecting the relative stability of the two states of the junctions.

At low currents, spin transfer torque effects can efficiently be modeled by an Arrhenius/N\'eel-Brown approach. The switching process is modeled by a Poisson process. If a junction is in the parallel state $P$, its probability to switch to the antiparallel state $AP$ during an infinitesimal time interval [$t$; $t+dt$) is $dt/\tau_P$. Conversely, if the junction is in the $AP$ state, its probability to switch to the $P$ state during the same interval is $dt/\tau_{AP}$. $\tau_P$ and $\tau_{AP}$ can be expressed as:

\begin{equation}
\tau_{P} = \tau_{0}\exp{ \left( \frac{\Delta E}{k T}\left(1-\frac{I}{I_{CP}} \right) \right) },
\label{tauP}
\end{equation} 
and
\begin{equation}
\tau_{AP} = \tau_{0}\exp{ \left( \frac{\Delta E}{k T}\left(1+\frac{I}{I_{CAP}} \right) \right) },
\label{tauAP}
\end{equation} 

where $\tau_{0}$ is a constant whose value is often assumed to be around $1~\mathrm{ns}$, $\Delta E$ is the energy barrier separating the two states of the junction, and $k T$ is the thermal energy.  The critical current $I_{CP}$ and $I_{CAP}$ can be expressed as

\begin{equation}
I_{CP/AP}= \frac{2e}{\hbar}\frac{1 \pm P^2}{P}\alpha V \mu_0
M_S \left(H_K+\frac{M_S}{2}\right),
\label{IC_IP}
\end{equation} 
for an in-plane magnetization junction, and
\begin{equation}
I_{CP/AP}= \frac{2e}{\hbar}\frac{1 \pm P^2}{P}\alpha V \mu_0
M_S H_K,\label{IC_P}
\end{equation} 
for a perpendicular magnetization junction. $P$ is the spin polarization of the junction. It is connected to the tunnel magnetoresistance by $TMR=\frac{2P^2}{1-P^2}$. $V$ is the volume of the free layer of the junction, $\mu_0$ is the vacuum permittivity, $M_S$ is the magnetization of the free layer, $H_K$ the anisotropy, $e$ the elementary charge and $\hbar$ the normalized Planck constant.

\section{ Impact of Read Disturb Effects and Scaling Potential of the Approach }

In order to evaluate the impact of the read disturb effect on random bit generation by superparamagnetic junctions, we developed a compact model of the junctions implementing the equations of the previous section. At each time step, the model computes the switching probability of the junction, and draws a random number to decide if a switching event occurred. The model was coded in the \textit{VerilogA} description language to ensure compatibility with major circuit simulators, and uses only standard \textit{VerilogA} function. 

For the model to give reliable results, it is essential that the simulator chooses time steps that are much smaller than the $\tau_{AP}$ of all the junctions in the circuit in the $AP$ state, and the $\tau_{P}$ of  the junctions in the $P$ state. This can easily be ensured by a \textit{bound\_step} function call within the \textit{VerilogA} code of the compact model. For junctions to be uncorrelated with each other, the seed of random number generation is a parameter of the device model, which can be declared as a variable parameter in Monte Carlo simulations.

For measuring the state of the superparamagnetic tunnel junctions, we investigated the use of a precharge sense amplifier (PCSA), a circuit initially proposed for the readout of ST-MRAM \cite{zhao_high_2009}, and presented in Fig.~\ref{fig:PCSA}. This circuit is very fast, highly energy efficient, and should expose the devices to minimal read disturb effects. We designed a PCSA using the design kit of a 28~nm CMOS commercial technology, and the Cadence integrated circuit design tools. The PCSA reads the state of a superparamagnetic tunnel junction by comparison with a reference resistor whose value is chosen between the resistances of the superparamagnetic tunnel junction in the $AP$ and $P$ states.
Circuit simulations with the Cadence Spectre simulator confirm that the circuit can read the state of the junction with extremely low energy: $2.2~\mathrm{fJ}$ for a junction of resistance $10~k\Omega$. Energy consumption depends only weakly on the junction resistance: it is $2.0~\mathrm{fJ}$  for a $10~M\Omega$ junction and $2.5~\mathrm{fJ}$ for a $100~\Omega$ junction.

\begin{figure}[!t]
\centering
\includegraphics[width=\columnwidth]{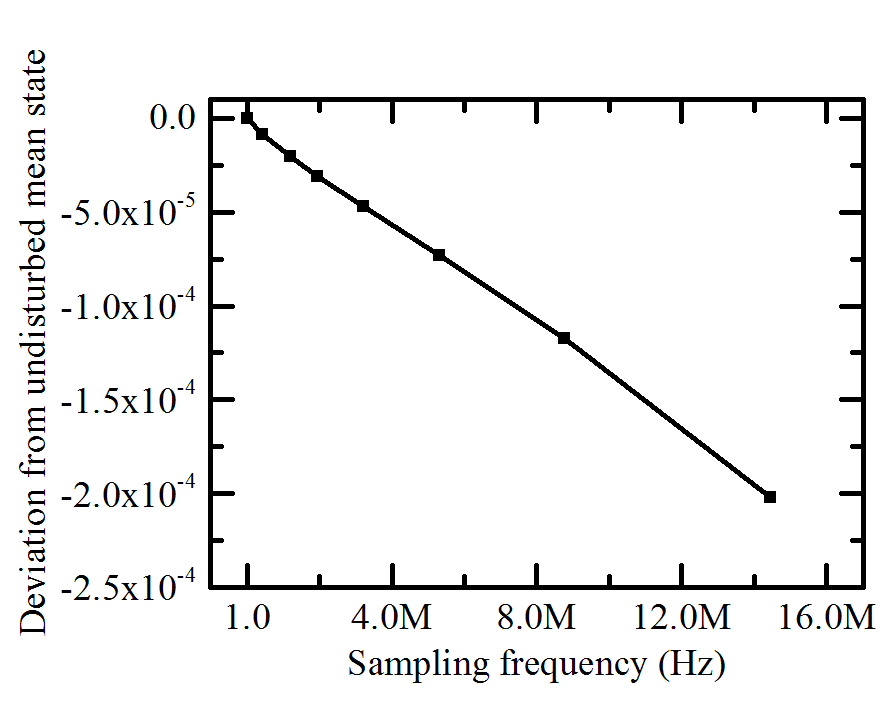}
\caption{Read disturb effect: deviation between the mean value read by a PCSA circuit $\bar{S}_\mathrm{PCSA}$ and the mean value of the superparamagnetic tunnel junction while not being read $\bar{S}_\mathrm{0}$. Results obtained by Monte Carlo simulations in Cadence Spectre, for junctions of different stability and speed (X axis).}
\label{fig:readdisturb}
\end{figure}

We then performed extensive Monte Carlo simulations with the Spectre simulator in order to extract the mean state measured by the PCSA. Fig.~\ref{fig:readdisturb} presents the relative deviation $(\bar{S}_\mathrm{PCSA}-\bar{S}_\mathrm{0})/\bar{S}_\mathrm{0}$ between mean state $\bar{S}_\mathrm{0}$ of the junction not subject to PCSA readout, and the mean measurement of the PCSA $\bar{S}_\mathrm{PCSA}$. This analysis was performed for junctions of different stabilities $\Delta E$. As the stability $\Delta E$ directly correlates with the natural frequency of the junction $\mathrm{F}_\mathrm{MTJ}$ through 
\begin{equation}
 \mathrm{F}_\mathrm{MTJ} = \frac{1}{\tau_{0}} \exp\left(-\frac{\Delta E}{k T}\right),
\label{fMTJ}
\end{equation} 
Fig.~\ref{fig:readdisturb} is presented as a function of $\mathrm{F}_\mathrm{MTJ}$.

For the most stable (slower) junctions, the impact of the read disturb effect appears minimal. It becomes more significant for less stable (faster) junctions and reaches approximately $0.01\%$ for junctions operating at $10~\mathrm{MHz}$.

Nevertheless, such deviations would not be a concern for applications as, for example, XOR8 whitening can compensate deviations as high as $10\%$, while ensuring that the results pass the NIST STS~\cite{PRA}. 
These considerations therefore confirm that a system associating PCSA readout and XOR8 whitening would be an appropriate way to extract high quality random bits from superparamagnetic tunnel junctions. Due to the simplicity of the circuit (PCSA, and XOR gates), its energy efficiency would be exceptional, requiring for example only 20~fJ/bit in the case of XOR8. This figure is orders of magnitudes below the CMOS state of the art of low energy truly random bit generation (3~pJ/bit in \cite{mathew_2.4_2012}). 

\section{Conclusion}
Superparamagnetic tunnel junctions are an attractive candidate for generating low energy random bits for alternative models of computation, as they rely on the same technology as ST-MRAM and naturally turn thermal noise into an easy to measure resistance state. In this work, we investigated the impact of associating a readout circuit with superparamagnetic tunnel junctions. We described the device modeling and showed, through circuit simulation, that even for very scaled junctions, read disturb is not a concern for applications if an efficient sense amplifier circuit is used for reading the state of the junction.

These results highlight that very low energy random bit generation is possible with superparamagnetic tunnel junctions,  $20~fJ/bit$ if XOR8 whitening is employed. The major limitation of the circuit is its relatively low speed, which might nevertheless be sufficient for many Internet-of-things applications envisioned by stochastic or cognitive computing.


\section*{Acknowledgment}

The authors would like to thank M.~Romera, A.~F.~Vincent, A.~Fukushima, K.~Yakushiji, H.~Kubota, S.~Yuasa, S.~Tiwari. This work is supported by the European Research Council Starting Grant NANOINFER (reference: 715872), and by the BAMBI EU collaborative FET Project grant (FP7-ICT-2013-C, project number 618024).



\bibliography{IEEEabrv,biblioconf}
\bibliographystyle{IEEEtran}

%



\end{document}